\documentclass[prd,preprint,nofootinbib,12pt]{revtex4}
\usepackage{amsfonts}
\usepackage{amsmath}
\usepackage{amssymb}
\usepackage{graphicx}
\usepackage{hyperref}
\usepackage{xcolor}

\newcommand{\dd}{{\rm{d}}} 

\newcommand{\im}{\mathrm{i}}

\begin{document}

\title{Comment on the ``New Rotating Black Hole in Electromagnetic Fields:
Cosmological Horizon without Cosmological Constant''}

\author{Hryhorii Ovcharenko}
\email{hryhorii.ovcharenko@matfyz.cuni.cz}
\affiliation{Charles University, Faculty of Mathematics and Physics,
Institute of Theoretical Physics,
V~Hole\v{s}ovi\v{c}k\'ach 2, 18000 Prague 8, Czechia}

\date{\today}

\begin{abstract}
    In this comment we discuss some properties of the novel spacetime, recently found in [L. Ma, H. L\"{u}, arXiv:2606.23782]. In particular, we draw attention to the background of this solution that the authors claim to be a new spacetime. We show that this is not the case because this background belongs to the special class of electrovacuum Kundt spacetimes of type D with either electric or magnetic charge. We show this by first analyzing the algebraic properties of this spacetime, and then by finding the explicit coordinate transformations. We hope that this analysis of the background allows for a better understanding of the structure of the general class of this type, namely as the Kerr-like black holes in the background generated by an accelerating electric or magnetic charge.
\end{abstract}

\maketitle

\newpage

\section{Introduction}

Investigation of exact solutions to the Einstein-Maxwell equations has always attracted significant attention. Among various non-trivial such exact spacetimes, we wish to mention a large class of black holes, immersed in external electromagnetic fields. In the way of solving this task, generating techniques appeared to be quite useful and productive. Half a century ago, Ernst and Wild \cite{Ernst1976,Ernst1976_2}, via employing the Harrison transformation, found a new class of spacetimes. It can be understood as
the Schwarzschild, Kerr, or Kerr-Newman black holes immersed in an external
magnetic field of the Bonnor-Melvin universe \cite{Bonnor1954,Melvin1964}. However, these methods mainly generate only black holes in the Bonnor-Melvin universe that has some drawbacks that do not allow it to be used as a realistic model of black holes in a magnetic field. Recently, a new class of spacetimes was found \cite{Ovcharenko2025} where authors without any use of these generating techniques, found a class of spacetimes that are immersed into another background, namely the Bertotti-Robinson one. This background represents the universe with a uniform electromagnetic field. In particular, it is worth noticing the Kerr-Bertotti-Robinson spacetime \cite{Podolsky2025} as a subclass of \cite{Ovcharenko2025} that is quite a simple solution, representing the Kerr spacetime in a uniform electromagnetic field, and it already finds wide astrophysical applications \cite{Vachher2025,Zhang2026,Wang2026,Li2026,Liu2025}.

Along with these studies, there appeared a branch of works \cite{Astorino2025,Astorino2026,Astorino2026_2,Barrientos2026,Barrientos2026_2,Ma2026_0} where the known generating techniques are used for the generation of new solutions, and as the seed used for such generation there are the solutions from the class presented in \cite{Ovcharenko2025}.

Recently, a new work in this branch of investigation by Ma and L\"{u} appeared \cite{Ma2026}, where the authors present a novel electrovacuum spacetime. This solution has a surprising property, namely that the non-trivial electromagnetic field in their solution may act like a positive cosmological constant. However, in this comment we draw attention to another aspect. Namely, for the case without mass and rotation, the authors obtained an electromagnetic background that they claim to be novel. We show that it is not the case, and that in fact this background belongs to the class of type D Kundt spacetimes with the non-trivial electromagnetic field. We hope that this comment may help with the interpretation of the whole class of solutions, presented in \cite{Ma2026}.

\section{The background properties and its explicit transformation to canonical form}

Let us start with a revision of the properties of the background solution, presented in \cite{Ma2026}. The metric is given by 
\begin{align}
    \dd s^2=\dfrac{1}{\big(1-\mathcal{E}^2 r^2(1-x^2)\big)^2}\Big[-(1-\mathcal{E}^2 r^2)\dd t^2+\dfrac{dr^2}{1-\mathcal{E}^2r^2}+r^2\dfrac{dx^2}{1-x^2}\Big]+r^2(1-x^2)\dd \varphi^2,\label{metr}
\end{align}
with the 4-potential
\begin{align}
    A_{(1)}^{ele}=-\dfrac{\mathcal{E}r x}{\sqrt{1-\mathcal{E}^2r^2(1-x^2)}}\dd t,~~~\mathrm{or}~~~A_{(1)}^{mag}=-\dfrac{\sqrt{1-\mathcal{E}^2r^2(1-x^2)}-1}{\mathcal{E}}\dd\varphi.\label{A_vec}
\end{align}
The authors claim that the background is new, as in this spacetime the relation
\begin{align}
    \mathrm{Riem}^2=14 (F^2)^2\label{spec_rel}
\end{align}
holds. As for both Bonnor-Melvin (BM) and Bertotti-Robinson (BR) spacetimes this condition does not hold, the authors claim that this spacetime is novel.

Let us inspect this statement. First of all, let us investigate the algebraic type of this spacetime. For the calculations, we use the natural null tetrad
\begin{align}
    \mathbf{k}=&\dfrac{\sqrt{f}}{\sqrt{2\Omega^2}}\Big(-\dd t+\dfrac{\dd r}{f}\Big),~~~\mathbf{l}=\dfrac{\sqrt{f}}{\sqrt{2\Omega^2}}\Big(-\dd t-\dfrac{\dd r}{f}\Big),\\
    \mathbf{m}=&\dfrac{r}{\sqrt{2(1-x^2)}}\Big(\im \dfrac{\dd x}{\Omega}-(1-x^2)\dd\varphi\Big),
\end{align}
where we introduced the shorthands
\begin{align}
    f=1-\mathcal{E}^2r^2,~~~\Omega=1-\mathcal{E}^2r^2(1-x^2).
\end{align}
The corresponding Weyl scalars with respect to this null tetrad are given by
\begin{align}
    \Psi_0&=\Psi_4=\dfrac{3}{2}\mathcal{E}^2(1-x^2)f\,\Omega,\\
    \Psi_1&=\Psi_3=\dfrac{3}{2}\mathcal{E}^2\im x\sqrt{1-x^2}\sqrt{f}\Omega,\\
    \Psi_2&=\dfrac{1}{2}\mathcal{E}^2\Omega (\Omega-3 x^2).
\end{align}

Calculating the invariants \cite{Stephani2003,Griffiths2009} for these Weyl scalars, one obtains
\begin{align}
    \Delta=I^3-17 J^2=0,~~~I\neq 0,~~~J\neq 0,~~~K=0,~~~M=0.
\end{align}
All these conditions mean that this spacetime is of algebraic type D. 

By constructing the null tetrad $(\mathbf{k}',\mathbf{l}',\mathbf{m}',\bar{\mathbf{m}}')$, aligned with the PNDs of the Weyl tensor, one indeed obtains that with respect to this tetrad:\footnote{We do not provide the explicit form of this tetrad here because it is quite complicated. It can be found in the supplementary material, see \cite{supp_mat}.}
\begin{align}
    \Psi_2'=&\mathcal{E}^2 \Omega^2,\\
    \Psi_0'=&\Psi_1'=\Psi_3'=\Psi_4'=0.
\end{align}

As the electromagnetic field is non-trivial for this background, we also have to investigate how it is aligned with the PNDs of the Weyl tensor. Calculating the NP scalars, one obtains:
\begin{align}
    \Phi_0'&=\Phi_2'=0,\\
    \Phi_1'&=-\dfrac{\im}{2}\mathcal{E}\,\Omega.
\end{align}
We see that the electromagnetic field is aligned with the PNDs of the Weyl tensor. To summarize, the spacetime (\ref{metr})-(\ref{A_vec}) 
\begin{enumerate}
    \item Is the electrovacuum spacetime,
    \item Is of algebraic type D,
    \item Has the aligned electromagnetic field.
\end{enumerate}
All spacetimes satisfying the conditions 1)-3) are known for a long time, and they were found by Debever, Kamran and McLenaghan in \cite{Debever1984}. In this work, the authors presented the most general spacetime satisfying the conditions 1)-3). Our current task is to understand which subclass of the Debever-Kamran-McLenaghan spacetime is the spacetime (\ref{metr})-(\ref{A_vec}). For this, let us inspect the optical scalars with respect to the PND-adapted null tetrad $(\mathbf{k}',\mathbf{l}',\mathbf{m}',\bar{\mathbf{m}}')$. By explicit calculations, one obtains that
\begin{align}
    \kappa&=\sigma=\nu=\lambda=\rho=\mu=0.
\end{align}

This means that both PNDs are geodesic, shear-free, expansion-free and non-twisting. These are important facts, as they mean that the space-time (\ref{metr})-(\ref{A_vec}) belongs to the Kundt subclass. All Kundt type D (electro)vacuum spacetimes were recently extensively investigated in \cite{Podolsky2018}. They are the generalizations of the B-metric \cite{Ehlers1962,Stephani2003,Griffiths2009} and in some cases can be interpreted as the gravitational field of a tachyon \cite{Gott1974,Hruska2019}. In particular, all the spacetimes (without the cosmological constant) are given by (see eqs. (22-23) from \cite{Podolsky2018})
\begin{align}
    \dd s^2=\rho^2\Big(-Q\,\dd \tau^2+\dfrac{\dd q^2}{Q}\Big)+\dfrac{P}{\rho^2}(\dd y+2 \gamma q\, \dd t)^2+\dfrac{\rho^2}{P}\dd p^2,\label{metr_Kundt}
\end{align}
with 
\begin{align}
    \rho^2&=\gamma^2+p^2,\\
    Q&=\epsilon_0-\epsilon_2 q^2,\\
    P&=\epsilon_2 p^2+2n p-(q_e^2+q_g^2)-\epsilon_2 \gamma^2.
\end{align}
where $q_e$ and $q_g$ are the electric and magnetic charges of this spacetime.
The electromagnetic field is given by the 4-potential.\footnote {This expression for the 4-potential is not explicitly given in \cite{Podolsky2018}, but it can be found by integrating the components of the Faraday tensor presented in Sec. 10.2 in \cite{Podolsky2018}, namely eq. (143).}
\begin{align}
    \mathbf{A}=-q\, \dfrac{q_e (\gamma^2-p^2)+2 q_g \gamma \,p}{\gamma^2+p^2}\dd t-\dfrac{p q_g+\gamma q_e}{\gamma^2+p^2}\dd \varphi.
\end{align}
In general, this spacetime is described by 6 parameters, namely ($\epsilon_0,\epsilon_2,n,\gamma,q_e,q_g$). However, we need to understand which set of these parameters corresponds to the spacetime (\ref{metr})-(\ref{A_vec}). For this, the condition (\ref{spec_rel}) appears to be extremely useful, and by inspecting the electromagnetic field for the Kundt type D spacetime, one obtains that (\ref{spec_rel}) is satisfied only in the following 3 cases
\begin{enumerate}
    \item $\gamma=0,~~~n=0,~~~q_g=0$,
    \item $\gamma=0,~~~n=0,~~~q_e=0$,
    \item $\epsilon_2=0,~~~n=0,~~~q_e=0=q_g$.
\end{enumerate}

However, in the third case the electromagnetic field vanishes, so it cannot correspond to the spacetime (\ref{metr})-(\ref{A_vec}) with a non-trivial electromagnetic field. Also in this case, the function $P$ vanishes, making the metric degenerate. Thus we are left only with the two possible cases: 1) and 2), with the metric
\begin{align}
    \dd s^2=p^2\Big(-Q\, \dd \tau^2+\dfrac{\dd q^2}{Q}\Big)+\Big(\epsilon_2-\dfrac{q_e^2+q_g^2}{p^2}\Big)\dd y^2+\Big(\epsilon_2-\dfrac{q_e^2+q_g^2}{p^2}\Big)^{-1}\dd p^2,
\end{align}
where
\begin{align}
    Q=\epsilon_0-\epsilon_2 q^2.
\end{align}

Indeed, if we perform the transformations of the coordinates:
\begin{align}
    y=\varphi/\mathcal{E},~~~\tau=\dfrac{\mathcal{E}}{\sqrt{\epsilon_0}}t,~~~q=\dfrac{\mathcal{E}\sqrt{\epsilon_0}r x}{\sqrt{1-\mathcal{E}^2r^2(1-x^2)}},~~~p=\dfrac{1}{\sqrt{1-\mathcal{E}^2r^2(1-x^2)}},
\end{align}
with $\epsilon_2=1$ and $q_e=-1/\mathcal{E}$ in the first case, and $q_g=1/\mathcal{E}$ in the second case, we obtain the metric (\ref{metr})-(\ref{A_vec}). The electrically charged background corresponds to the case 1), while the magnetically charged to the case 2).

Finally, we wish to comment on the interpretation of the Kundt spacetime (\ref{metr_Kundt}). The whole class of spacetimes (\ref{metr_Kundt}) was first found by Carter in \cite{Carter1968}, and it was nicknamed as the $[\tilde{B}(-)]$ metric. Also, this spacetime was rediscovered by Pleba\'{n}ski \cite{Plebanski1975} and Pleba\'{n}ski-Demia\'{n}ski \cite{Plebanski1976}. Here we deal with the case of zero $\gamma$ and $n$ parameters, because only in these cases there exists a transformation to (\ref{metr})-(\ref{A_vec}) (but only when there is either an electric $q_e$ or a magnetic $q_g$ charge). The meaning of this Kundt spacetime can be understood if we first consider the chargeless limit. In this case, as was shown in Sec. 5 in \cite{Podolsky2018}, this spacetime is Minkowski background, but expressed in accelerating coordinates. Thus, the interpretation of the spacetime (\ref{metr})-(\ref{A_vec}) is the \textit{metric of an accelerating (electric or magnetic) charge}. 

This allows one to interpret the whole spacetime, found in \cite{Ma2026}, as the Kerr black hole located in the background, created by accelerating electric (in the first case) or magnetic (in the second case) charge.

\section{Acknowledgements}
This work was supported by the Czech Science Foundation Grant No. GA\v{C}R 26-22381S,
and by the Charles University Grant No. GAUK 260325. The author thanks Ji\v{r}\'{i} Podolsk\'{y} and David Kubiz\v{n}\'{a}k for valuable discussions and comments.

\end{document}